\documentclass{vgtc}                          
\let\ifpdf\relax

\usepackage{mystyle}
\usepackage{subfigure}
\usepackage{algorithm}
\usepackage[noend]{algorithmic}
\usepackage{boxedminipage}
\usepackage{review}
\usepackage{url}

\usepackage{booktabs}

\newcommand{\techRS}{\textsf{RS}\xspace}
\newcommand{\techZ}{\textsf{Z-order}\xspace}
\newcommand{\etal}{et{.}al{.}}

\usepackage{mathptmx}
\usepackage{graphicx}
\usepackage{times}
\usepackage{booktabs}

\onlineid{0}

\vgtccategory{Research}

\vgtcinsertpkg

\title{Visualization of Big Spatial Data using \\ Coresets for Kernel Density Estimates}

\author{Yan Zheng\thanks{e-mail: yanzh.cs@gmail.com; Much of this work was completed while at the University of Utah.}\\ %
        \scriptsize Visa Research %
 \and  Yi Ou\thanks{e-mail: olly93219@outlook.com; Much of this work was completed while at the University of Utah.}\\ %
        \scriptsize Expedia, Inc. %
\and  Alexander Lex\thanks{e-mail: alex@sci.utah.edu}\\ %
        \scriptsize  University of Utah 
\and Jeff M. Phillips\thanks{e-mail: jeffp@cs.utah.edu }\\ %
     \scriptsize University of Utah %
}

\abstract{
The size of large, geo-located datasets has reached scales where visualization of all data points is inefficient. Random sampling is a method to reduce the size of a dataset, yet it can introduce unwanted errors.	We describe a method for subsampling of spatial data suitable for creating kernel density estimates from very large data and demonstrate that it results in less error than random sampling. We also introduce a method to ensure that thresholding of low values based on sampled data does not omit any regions above the desired threshold when working with sampled data. We demonstrate the effectiveness of our approach using both, artificial and real-world large geospatial datasets.
} 


\begin{document}



\maketitle

\section{Introduction} 

Data is collected at ever-increasing sizes, and for many datasets, each data point has geo-spatial locations (e.g., either (x,y)-coordinates, or latitudes and longitudes).  
Examples include population tracking data, geo-located social media contributions, seismic data, crime data, and weather station data. 
The availability of such detailed datasets enables analysts to ask more complex and specific questions. These have applications in wide ranging areas including 
biosurveillance, 
epidemiology, 
economics, 
ecology environmental management, 
public policy and safety, 
transportation design and monitoring, 
geology, and 
climatology. 
Truly large datasets, however, cannot be simply plotted, since they typically exceed the number of pixels available for plotting, the available storage space, and/or the available bandwidth necessary to transfer the data.  

\begin{figure}[t!]

\vspace{-.21in}

 \begin{center}
 	\hspace{0.02\linewidth}
 \includegraphics[width=0.43\linewidth]{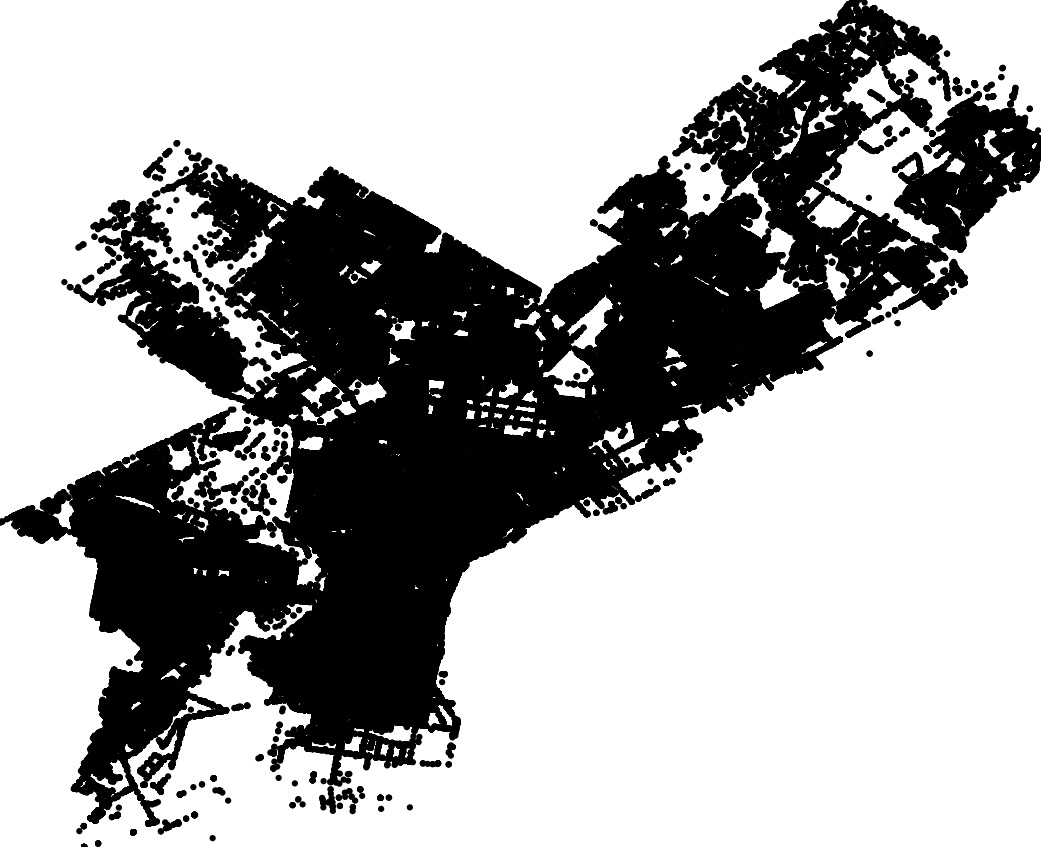}
 \hfill
 \includegraphics[width=0.43\linewidth]{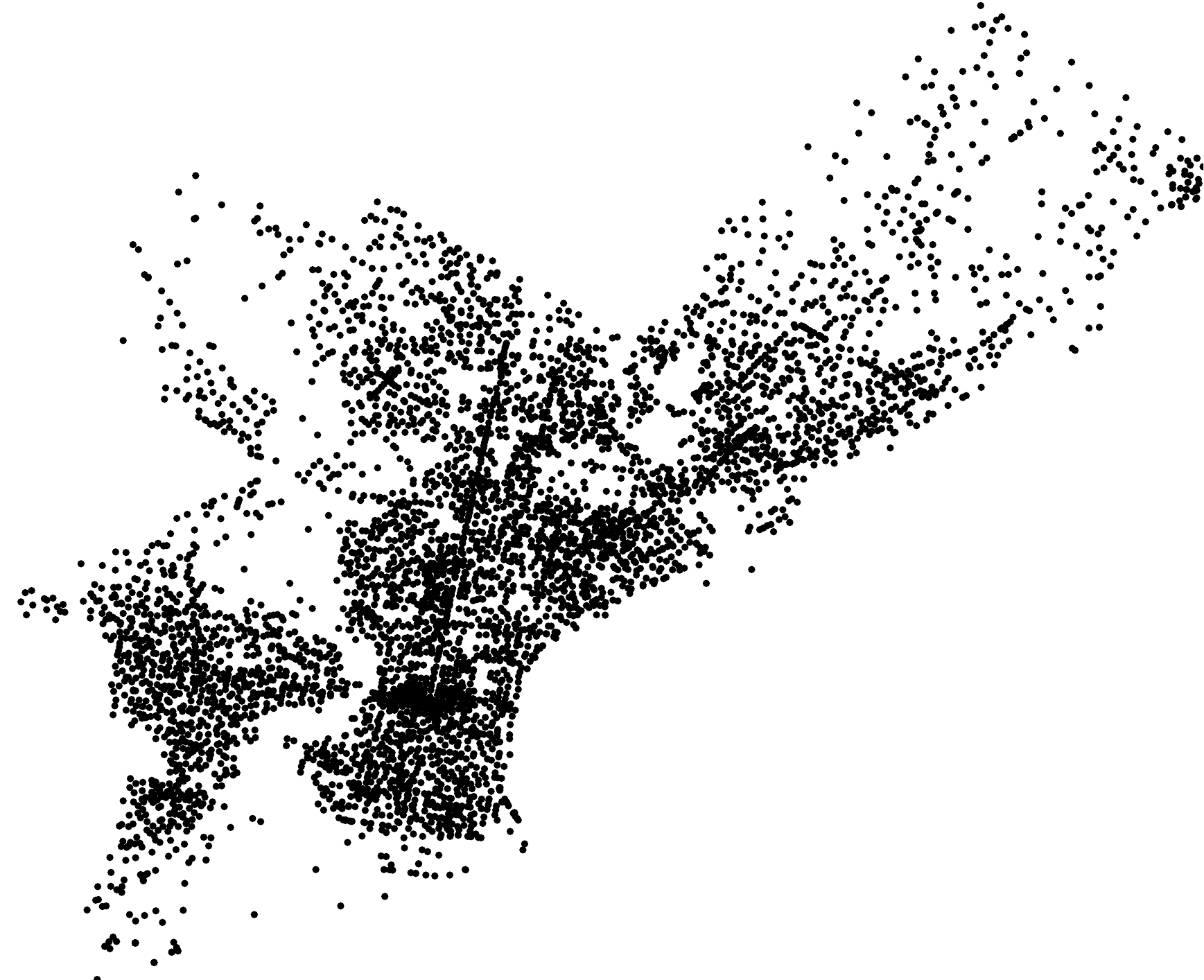}
  	\hspace{0.02\linewidth}
 \end{center}
 \vspace{-5mm}
  \caption{   \label{fig:orgvscore}
Crimes from 2006 to 2013 in Philadelphia, the full dataset (left) with 0.7 million points and a coreset (right) with only 5300 points.}
 \vspace{-7mm}
\end{figure}

A common way to manage and visualize such large, complex spatial data is to aggregate it using a kernel density estimate~\cite{Sil86,Sco92} ($\kde$).  
A $\kde$ is a statistically and spatially robust method to represent a continuous density using only a discrete set of sample points.  Informally, this can be thought of as a continuous average over all choices of histograms, which avoid some instability issues that arise in histograms due to discretization boundaries.  
For a formal definition, we first require a kernel $K : \b{R}^2 \times \b{R}^2 \to \b{R}$; we will use the Gaussian kernel $K(p,x) = e^{-\|p-x\|^2}$.  
Then, given a planar point set $P \subset \b{R}^2$, the kernel density estimate is defined at any query point $x \in \b{R}^2$ as
\[
\kde_P(x) = \frac{1}{|P|} \sum_{p \in P} K(p,x).
\]
This allows regions with more points nearby (i.e., points $x$ with a large value $K(p,x)$ for many $p$ in $P$) to have a large density value, and this function is smooth and in general nicely behaved in several contexts.  Using this function summarizes the data, and avoids the over-plotting and obfuscation issues demonstrated in Figure \ref{fig:orgvscore}(left).  
However, just computing $\kde_P(x)$ for a single value $x$  requires $O(|P|)$ time.  
While these values can be precomputed and mapped to a bitmap, visually interacting with a $\kde$ e.g., to query and filter, would then require expensive reaggregating.

Towards alleviating these issues, we propose to use \textbf{coresets} for $\kde$s.  
In general, a \emph{coreset} $Q$ is a carefully designed small subset of a very large dataset $P$ where $Q$ retains properties from $P$ as accurately as possible.  In particular, in many cases the size of $Q$ depends only on a desired minimum level of accuracy, not the size of the original dataset $P$.  This implies that even if the full dataset grows, the size of the coreset required to represent a phenomenon stays fixed.  This also holds when $P$ represents a continuous quantity (like the locus of points along a road network) and $Q$ constitutes some carefully placed way-points~\cite{SFR12}.  Figure \ref{fig:orgvscore} shows a dataset $P$ with 700 thousand points and its coreset from all reported crimes in Philadelphia from 2005-2014.  
For more details on variations and constructions, refer to recent surveys~\cite{Phi16,BLK17}.  

%
%

In particular, a coreset for a kernel density estimate is a subset $Q \subset P$~\cite{Phillips2013,big-kde}, with $|Q| \ll |P|$  so that for some error parameter~$\eps$ 
\begin{equation}\label{eq:Linfty}
L_\infty(\kde_P,\kde_Q) = \max_{x \in \b{R}^2} | \kde_P(x) - \kde_Q(x)| \leq \eps.
\end{equation}
This means that at any and all evaluation points $x$, the kernel density estimates are guaranteed to be close.  In particular, such a bound on the \emph{worst case error} is essential when attempting to find outlier or anomalous regions; in contrast an average case error bound (e.g. $L_1(\kde_P, \kde_Q)$) would allow for false positives and false negatives even with small overall error.  Thus, with such a worst-case bounded coreset $Q$, we can use $\kde_Q$ efficiently without misrepresenting the data.  

In the rest of this paper we demonstrate two properties of coresets used for $\kde$s that make them pertinent for visual analysis. 
In Section \ref{sec:coresets}, we first demonstrate that we can create a coreset that is more accurate than the naive but common approach of random sampling.  
Second, very sparse subsets (e.g., from random sampling) tend to cause anomalous regions of low, but noticeable density; we introduce a method to counteract this problem in Section \ref{sec:enet}, by carefully adjusting the smallest non-zero layer of the corresponding transfer function.  
Towards demonstrating these insights we design and present an interactive system for visualizing large, complex spatial data with coresets of kernel density estimates.  
Based on these insights, we believe that coresets and kernel density estimates can become an important tool for interactive visual analysis of large spatial data.  


\section{Related Work}
\label{sec:related}

Visualizing large spatial datasets is a challenge attracting a lot of attention among the visualization community.  This has led to the development of a variety of research platforms including Polaris~\cite{STH02}, inMens~\cite{LJH13}, Nanocubes~\cite{nanocubes} and Gaussian cubes~\cite{ZFWBS16}.  These systems all provide a variety of ways of to explore, interact, and analyze spatial datasets.  
For interacting with such spatial data purely based on its density, a kernel density estimate is a necessary and often the default tool; it is the statistical premise behind a heat map.  

Another common theme among visualization systems for large data is that in order to allow 
real-time interaction, every single data point cannot be rendered.  The data somehow needs to be compressed, either as a subset, or by some statistical summarization.  
This trend denominates  efficiency and scalability focused database projects, such as BlinkDB~\cite{BlinkDB} and STORM~\cite{STORM}.  In these systems, random sampling of data is the core tool since it can be done efficiently and preserves to some degree most relevant statistical properties of the data.  

\begin{figure}[h]
\label{fig:STORM}
\centering \includegraphics[width=0.7\linewidth]{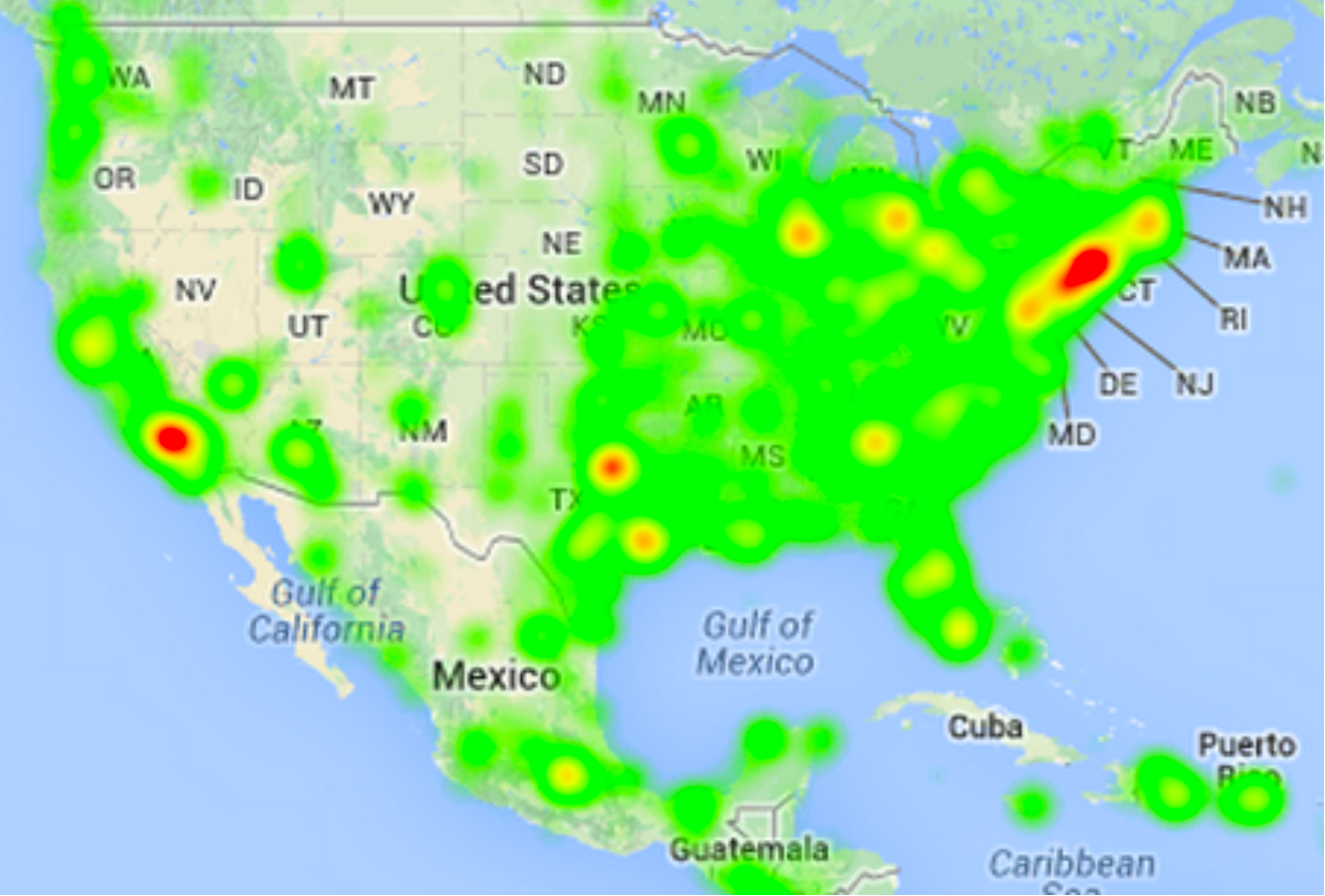}
\caption{Screenshot image from STORM~\cite{STORM} showing heatmap/KDE of tweet density in the USA.}
\end{figure}

Numerous other sampling schemes have been proposed to reduce the dataset size for visualization~\cite{PCM16,KBPIMR15}.  However, these approaches do not directly address the preservation of kernel density estimates.  Park \etal~\cite{PCM16} develop heuristics to optimize a measure related to the inverse of a \kde, and consider mainly data from long strands of data along road networks.  Kim \etal~\cite{KBPIMR15} focus on techniques for binned, one-dimensional data.  Moreover, these approaches are considerably more complicated than the ones we consider and do not allow for efficient and stable updates in the parameters of the \kde.

\section{Coresets Constructions}
\label{sec:coresets}

When tracking tweets or when analyzing crime in an area, a high frequency of such events in a sparsely populated area can be an important pattern to analyze further.
If a subset has low error on average, but has locations with large deviations from the truth, analysis based on that subset can lead to both false positives and negatives.  This is why the $L_\infty$ error, as in equation (\ref{eq:Linfty}) is the right way to measure accuracy. Both coresets techniques for $\kde$s~\cite{big-kde} and random sampling can make such guarantees, but the ones for coresets are stronger.

\begin{enumerate}
\item A random sample $Q$ of size $O((1/\eps^2) \log(1/\delta))$ from a large set $P$ creates a coreset for kernel density estimates with probability at least $1-\delta$~\cite{JoshiKommarajuPhillips2011}.  We refer such a method as \techRS.  This can be implemented in $O(|P|)$ time.  

\item There are several techniques to create coresets for kernel density estimates~\cite{JoshiKommarajuPhillips2011,big-kde,Phillips2013,CWS10}.  The one we use~\cite{big-kde} (labeled \techZ, described below) results in a coreset of size $O((1/\eps) \log^{2.5} (1/\eps) \log(1/\delta))$, that succeeds with probability at least $1-\delta$, and runs in time $O(|P| \log |P|)$ time.  This is roughly a square-root of the size of the random sample technique.  Note that other techniques~\cite{Phillips2013}, can in theory reduce the coreset size to $O((1/\eps) \log^{0.5} (1/\eps))$; the \techZ method mimics this approach with something more efficient and with better constant factors, but a bit worse ``in theory.''  
\end{enumerate}

\begin{figure}[h!]
\vspace{-.15in}

	\hspace{0.1\linewidth}
	\begin{minipage}[t]{0.15\linewidth}  
		\vspace{-20mm}
		{\small
			\begin{tabular}{ccc}
				Size & RS Err & Coreset Err\\
				\toprule
				830 &  0.035 & 0.01\\
				1890 & 0.023 & 0.005\\
				5000 & 0.014 & 0.002\\
				10000 & 0.01 & 0.001\\
			\end{tabular}
		}
	\end{minipage}
	\hfill
	\begin{minipage}[t]{0.5\linewidth}
		\centering
		\includegraphics[width=1.3in]{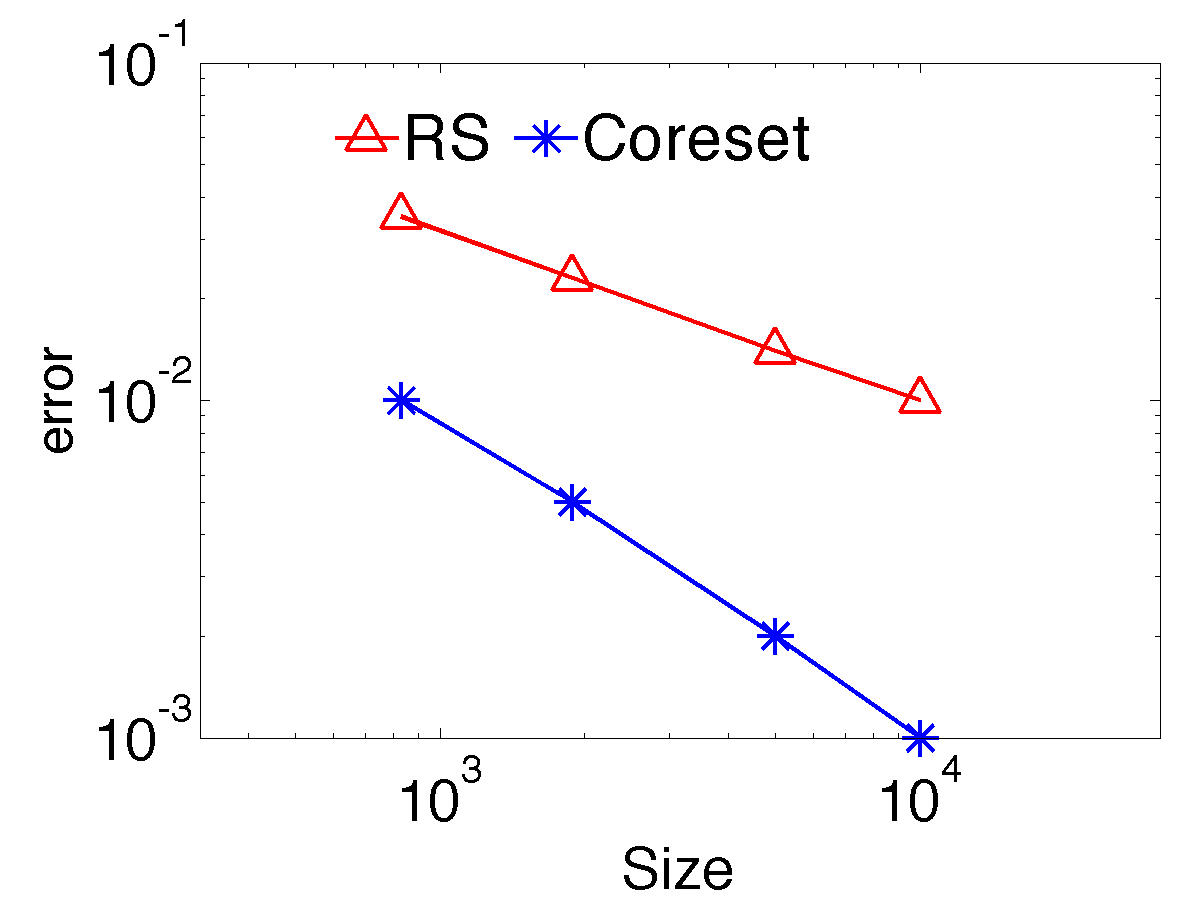}  
	\end{minipage}
	\vspace{-3mm}
	\caption{Error comparison of random sample (RS) and coresets. }
	\vspace{-5mm}
	\label{fig:errsize}
\end{figure}

While these theoretical bounds are useful guidance for effectiveness of these techniques, we also demonstrate them empirically in Figure \ref{fig:errsize} using Open Street Map Utah highway data.  We observer that indeed \techZ produces a coreset roughly a square-root of the size of the one produced by \techRS for the same observed error.  

\subsection{Coreset method}
To generate the coresets, we use the two-dimensional technique based on space filling curves~\cite{big-kde}. A space filling curve~\cite{ARRWW97} puts a single order on two- (or higher-) dimensional points that preserves spatial locality.  They have many uses in databases for approximate high-dimensional nearest-neighbor queries and range queries.  The single order can be used for a (one-dimensional) B$^{+}$-tree, which provides extremely efficient queries even on massive datasets that do not fit in memory.

In particular, the \techZ curve is a specific type of space filling
curve that can be interpreted as implicitly ordering points based on
the traversal order of a quad tree.  That is if all of the points are
in the range $[0,1]^2$ (or normalized to be so), then the top level of the quad tree has $4$ children
over the domains $c_1 = [0,\frac{1}{2}] \times [0,\frac{1}{2}]$, $c_2=
[\frac{1}{2},1] \times [0,\frac{1}{2}]$, $c_3=[0,\frac{1}{2}] \times
[\frac{1}{2},1] $, and $c_4= [\frac{1}{2},1] \times [\frac{1}{2},1]$.
Each child's four children itself divide symmetrically, and so on recursively.
Then the \techZ curve visits all points in the child $c_1$, then all
points in $c_2$, then all points in $c_3$, and all points in $c_4$ (in
the shape of a`Z'); and all points within each child are
also visited in such a Z-shaped order.  
Thus given a domain containing all points, this
defines a complete order on them, and the order generally preserves spatial
locality as well as a quad tree does.  Usefully, the order of two points can be directly compared without knowing all of the data, so plugging in such a comparison operation, any efficient comparison-based sorting algorithm can be used to sort points in this order.  

To generate the coreset based on the \techZ curve, set $k=O(\frac{1}{\eps}
\log^{2.5}\frac{1}{\eps})$ and randomly select one point from each
\techZ rank range $[(i-1)\frac{|P|}{k}, i\frac{|P|}{k}]$.  The resulting set $Q$ gives an $\eps$-sample of $\kde$.
Note that this approach is oblivious to the parameters in the kernel density estimate (the type of kernel, the choice of bandwidth, the bitmap on which it is visualized), so it does not need to be updated if we change these parameters.

\subsection{Pre-ordering points}
One downside of the above method, is that if we would like to change the resolution of the coreset, that is increase or decrease its accuracy by increasing or decreasing its size, we need to repeat much of the computation.  Sorting the $|P|$ points takes $O(|P| \log |P|)$ time, and selecting a coreset from the sorted list would take $O(|P|)$ time under most implementations and ways of preprocessing the data.

\begin{table*}[t]
\begin{center}
\begin{tabular}{r  c  c  c  c  c   c  c  c }

\textbf{Input: \techZ index}  & \textbf{1} & \textbf{2} & \textbf{3} & \textbf{4} & \textbf{5} & \textbf{6} & \textbf{7} & \textbf{\texttt{x}}
\\ 
\midrule
binary representation & 000 & 001 & 010 & 011 & 100 & 101 & 110 & 111
\\
reverse bits & 000 & 100 & 010 & 110 & 001 & 101 & 011 & 111
\\
after random mask $M=101$ & 101 & 001 & 111 & 011 & 100 & 000 & 110 & 010
\\
new binary ordering index & 6 & 2 & 8 & 4 & 5 & 1 & 7 & 3
\\
priority ordering index & 5 & 2 & 7 & 3 & 4 & 1 & 6 & \texttt{x}
\\

\end{tabular}
\end{center}

\vspace{-.2in}
\caption{\label{tbl:example-reorder}
An example demonstration of using bit-reversal to create a priority ordering.  The first line describes the input Zordering index, based on this sorted order.  There are $7$ points and one dummy point designated as \texttt{x}.  The final line indicates the resulting priority ordering after removing the dummy point.}
\end{table*}

Rather we propose a more useful way to preprocess the data.  In particular, we can reorder the original dataset $P$ (from the \techZ to a different ordering) to what we call a \emph{priority ordering}, so that the first $k$ points in that order are precisely the points to choose as a coreset of size $k$. For instance, such a priority ordering can be created via random sampling: assign each point a random number, and sort on the points by these random numbers. This priority ordering has several enticing properties.  
\begin{itemize}
\item The coreset construction only needs to be done once, and this can be done offline and in code that lives outside of an interactive visualization system.  For instance, in our implementation, this is realized extremely efficiently in low-level C, but we have built our visualization in JavaScript, Canvas, and D3. This also makes the visualization system modular, separating the coreset construction technique, which only needs to provide a (priority) ordered set of points.  
\item If we increase the size of the coreset, the new larger coreset necessarily contains the old smaller one.  This increases the stability of the result, since for instance increasing the size $k$ by one point, only changes the coreset by $1$ point.  This means adjusting this parameter makes the visual interface more efficient and less jarring.  Also, for small updates, it can allow for some caching in recomputing various quantities.  
In contrast, for a coreset $Q_1$ constructed directly from a \techZ, if the size parameter is changed slightly, we may recompute a new coreset $Q_2$ to satisfy this parameter change with no overlap with $Q_1$.  This could cause the visualization to appear unstable and require that everything is completely recomputed.  
\end{itemize}

For the \techZ approach, we can simply describe this priority reordering using a bit reversal.  Given all of the points sorted by the \techZ, label each point as a binary number starting from $0\ldots00$, $0\ldots01$, $0\ldots10$, $0\ldots11$, $\ldots$.  Pad the dataset with dummy points so the total number is a power of $2$; i.e., all binary numbers of a fixed length are included.  Then reverse the order of the bits, so $101011$ becomes $110101$.  Next randomize this by taking a random mask $M$ and \textsc{xor}ing the mask with all flipped numbers; basically this randomly flips half of the bits.
Then sort these points by these new binary numbers.  Remove the dummy points, and this is the new order.   This is illustrated in a small example in Table \ref{tbl:example-reorder}.

\begin{figure}[t!]
  \includegraphics[width=0.98\linewidth]{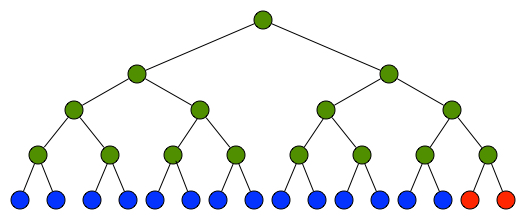}
  \caption{Index tree of a dataset of $14$ points (blue). Dummy nodes are shown in red.}
   \label{fig:indextree}
\end{figure}

An alternative way of understanding this approach is to illustrate it using a binary tree.  
For the original data $P$, we give each point an index $i$ based on the order of the points in the \techZ.  Then we construct a binary tree over these points based on this sorted order. Next, we fill up the binary tree with dummy points at the end of the ordering so that the size is a power of $2$, and the binary tree is a perfectly-balanced tree; see Figure \ref{fig:indextree} for an example with $14$ points.  

Then we re-order these points by selecting points from the tree in a random way, so the number of selected points in each subtree is as balanced as possible; Algorithm \ref{alg:sort} provides psuedocode for this priority re-ordering algorithm.
At each step, at each internal node, we keep track of how many points have been selected from each subtree.  If the two subtrees have the same number of selected points, choose one at random, and recurse.  If the two subtrees have imbalanced counts of selected points, then recurse on the subtree (which will be \texttt{unmarked}) that has fewer selected points.  This randomizes the process while ensuring that the selection is as balanced as possible with respect to the original ordering.  
The new, priority order of the points $S = \langle s_1, s_2, \ldots, s_n\rangle$ is the order in which they are selected, ignoring dummy points. The purpose of the dummy points it to make sure that we don't over-select from the existing points on the right subtree if they have fewer points than the left subtree.

\begin{algorithm}
  \caption{\textbf{priority reordering}}
  \label{alg:sort}
    \begin{algorithmic}[1]
    \STATE $i=1$
    \LOOP
    \STATE node = root
    \WHILE{(node is not leaf node \emph{and} not \texttt{marked})}
     \IF {(node$\rightarrow$left \emph{and} node$\rightarrow$right are both \texttt{unmarked})}
    	\STATE generate a random number $r$ from $\{0,1\}$
        \STATE \textbf{if} $r = 0$ \textbf{then} node = node$\rightarrow$left \emph{and} \texttt{mark} node
        \STATE \textbf{if} $r=1$ \textbf{then} node = node$\rightarrow$right \emph{and} \texttt{mark} node
     \ELSIF{(node$\rightarrow$left is \texttt{marked})}
         \STATE reset node$\rightarrow$left as \texttt{unmarked} 
         \STATE node = node$\rightarrow$right 
     \ELSIF{(node$\rightarrow$right is marked)}
         \STATE reset node$\rightarrow$right as \texttt{unmarked} 
         \STATE node = node$\rightarrow$left 
     \ELSIF{(both children are \texttt{marked})} 
     	  \STATE \textbf{return}  \hspace{.3in}  \emph{[all nodes have been processed]} 
     \ENDIF
	\IF {(leaf node \emph{and} not dummy)}
	    \STATE output node as $s_i$
	    \STATE $i = i+1$
	\ENDIF
     \ENDWHILE
    \ENDLOOP
   \end{algorithmic}
 \vskip1pt
\end{algorithm}

\begin{figure*}[h!]

 	\hspace{25mm} \textsf{Full Data}   \hspace{45mm}  \textsf{Coreset}  \hspace{40mm}  \textsf{Random Sample} \\
 	 	\rotatebox{90}{\hspace{7mm} \textsf{Kentucky }} 
 	\fbox{\includegraphics[width=0.30\linewidth]{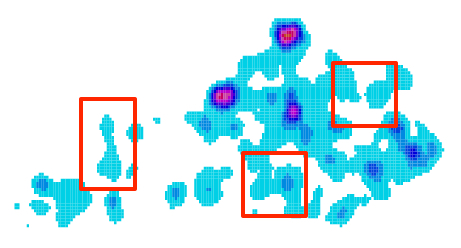}}
 	\fbox{\includegraphics[width=0.30\linewidth]{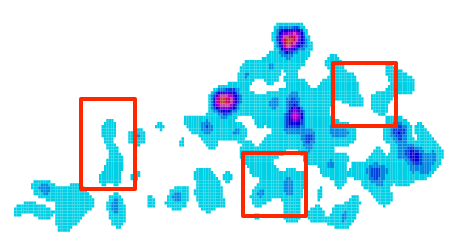}}
 	\fbox{\includegraphics[width=0.30\linewidth]{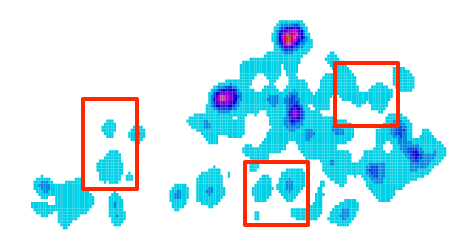}}
 	\rotatebox{90}{\hspace{15mm} \textsf{Philadelphia}} 
 	\fbox{\includegraphics[width=0.30\linewidth]{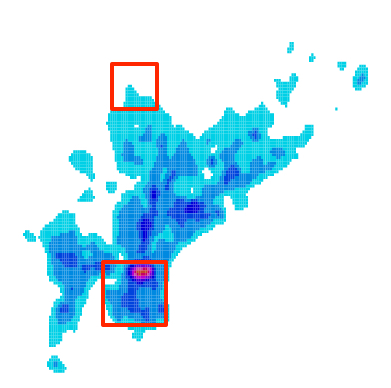}}
 	\fbox{\includegraphics[width=0.30\linewidth]{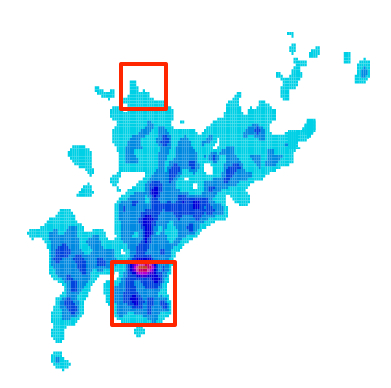}}
 	\fbox{\includegraphics[width=0.30\linewidth]{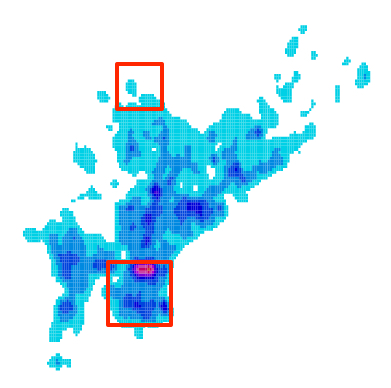}}
 	\vspace{2mm}
    \rotatebox{90}{\hspace{15mm} \textsf{Synthetic}} 
 	\fbox{\includegraphics[width=0.30\linewidth]{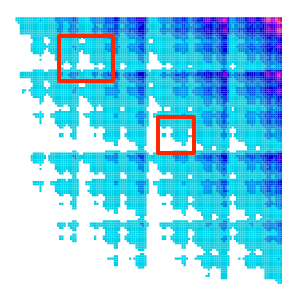}}
 	\fbox{\includegraphics[width=0.30\linewidth]{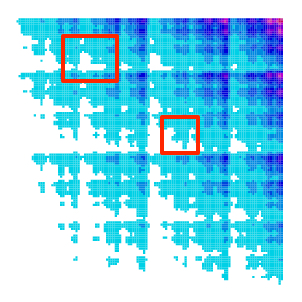}}
 	\fbox{\includegraphics[width=0.30\linewidth]{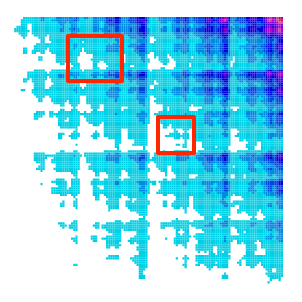}}
 	\hspace{30mm} \includegraphics[width=0.25\linewidth]{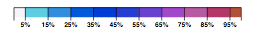}\\
 	\vspace{-2mm}
 	\caption{Comparison of ground truth $\kde$ (left), coreset $\kde$ (middle), random sample $\kde$ (right), on three datasets. 
 		Regions of high error in the random sampling are highlighted with red frames across all conditions.}
 	\label{fig:rsvscore}
 	\vspace{-5mm}
\end{figure*}

\subsection{Comparing Coresets with Random Sampling}

To guarantee $\eps$-error coresets require $O((1/\eps) \log^{2.5}\frac{1}{\eps})$ size, while random sampling requires $O(1/\eps^2)$ size.  In other words, coresets with the same error as random sampling can be about a square root of the size (see Figure~\ref{fig:errsize}).  We will compare two kind of errors: general error and relative error between original data $\kde$ and coreset $\kde$ as well as original data $\kde$ and random sample $\kde$. Suppose the original dataset is $P$,  coreset or random sample is defined as $Q$, then absolute error is defined as $\kde_P - \kde_Q$ and relative error is defined as $\frac{\kde_P - \kde_Q}{\kde_P}$.

\subsubsection{Datasets}
In our experiments we use two large real datasets and one synthetic dataset.  
The first dataset (\textit{Kentucky}) is of size $199{,}163$ and consists of the longitude and latitude of all highway data points from OpenStreetMap data in the state of Kentucky.   
The second dataset (\textit{Philadelphia}) contains $683{,}499$ geolocated data points; it consists of the longitude and latitude of all crime incidents reported in the city of Philadelphia by the Philadelphia Police Department between 2005 and 2014.  

Our \textit{Synthetic} dataset mimics a construction of Zheng and Phillips~\cite{zheng2015error} meant to create density features at many different scales using a recursive approach inside a unit square $[0,1]^2$. The dataset contains $532{,}900$ data points.
At the top level it generates $4$ points $p_1=(0,0), p_2=(0,1), p_3=(1,0), p_4=(1,1)$.  
We recurse into $9$ new rectangles by splitting the $x$- and $y$-coordinates into $3$ intervals each and taking the cross-product of these intervals.  The intervals are defined non-uniformly, splitting the $x$-range (and $y$-range) into pieces $[0,0.5]$, $[0.5, 0.8]$, and $[0.8,1.0]$.  We also add $4$ new points at $(0.5, 0.5)$, $(0.5, 0.8)$, $(0.8,0.5)$, and $(0.8, 0.8)$ to the created dataset. In recursing on the $9$ new rectangles we further split each of these and add points proportional to the length of their sides.  


\subsubsection{Visual Demonstration on Data}
To demonstrate the advantage of the coreset method over the random sampling method, we show the visualizations of $\kde$s on these three datasets in Figure \ref{fig:rsvscore}.  In this figure we show the $\kde$ of the original dataset, the coreset, and a random sample.  
We set the size of the coreset in \textsf{Kentucky} to $7{,}675$, in \textsf{Philadelphia} to $7{,}675$, and in \textsf{Synthetic} to $69{,}077$.  
A transfer function colors each pixel with respect to the largest $\kde(x)$ value observed in the full dataset (a dark red), transitioning to a light blue and then white for values less than $5\%$ of this value.  

The high-level structure for both the coreset and random sample visualizations are preserved in each case; however, for each dataset there are many subtle differences where the random sample captures some area incorrectly.  We have highlighted a few of these differences across the $3$ visualizations in red boxes in Figure \ref{fig:rsvscore}.  

\begin{figure*}[t]

 	\hspace{11mm} \textsf{$\kde_{\textsf{\small full}} - \kde_{\textbf{\small coreset}}$}   
	\hspace{15mm}  \textsf{$\kde_{\textsf{\small full}} - \kde_{\textbf{\small RS}}$}  
	\hspace{18mm}  \textsf{$\frac{\kde_{\textsf{\small full}} - \kde_{\textbf{\small coreset}}}{\kde_{\textsf{\small full}}}$} 
	\hspace{18mm}  \textsf{$\frac{\kde_{\textsf{\small full}} - \kde_{\textbf{\small RS}}}{\kde_{\textsf{\small full}}}$} \\ 
 	\rotatebox{90}{\hspace{7mm} \textsf{Kentucky}} 
 	\fbox{\includegraphics[width=0.22\linewidth]{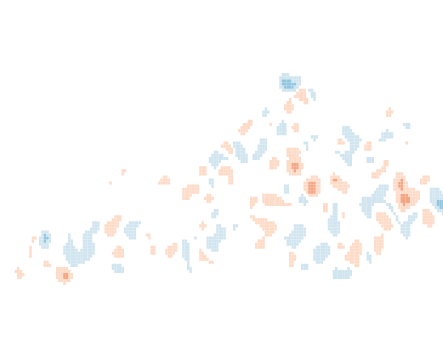}}
 	\fbox{\includegraphics[width=0.22\linewidth]{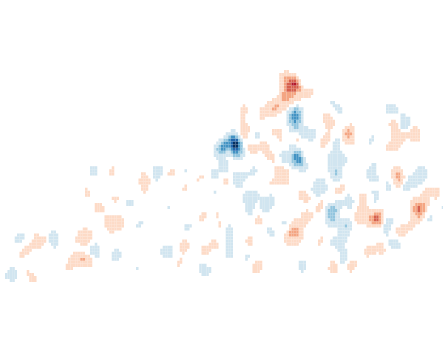}}
 	\fbox{\includegraphics[width=0.22\linewidth]{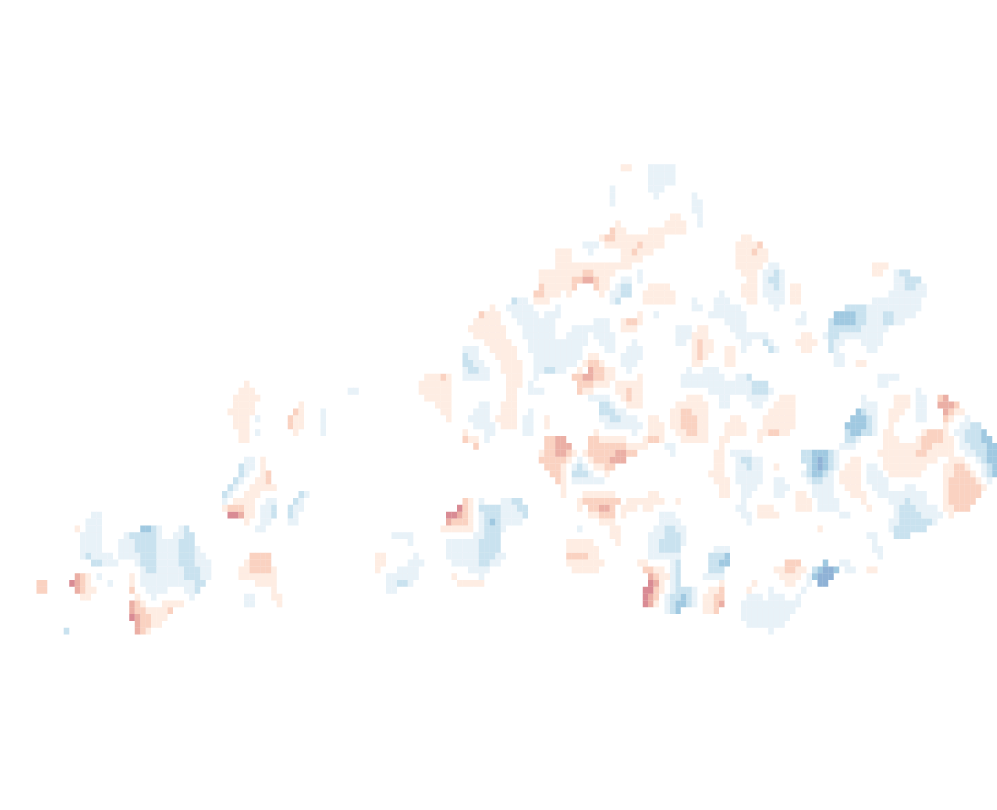}}
 	\fbox{\includegraphics[width=0.22\linewidth]{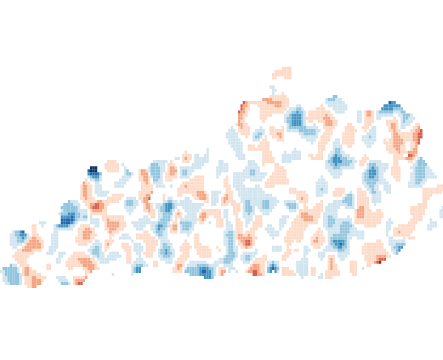}}
 	\vspace{2mm}
 	\\
 	\rotatebox{90}{\hspace{7mm} \textsf{Philadelphia }} 
 	\fbox{\includegraphics[width=0.22\linewidth]{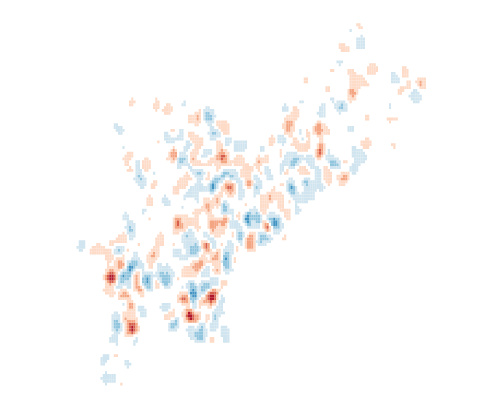}}
 	\fbox{\includegraphics[width=0.22\linewidth]{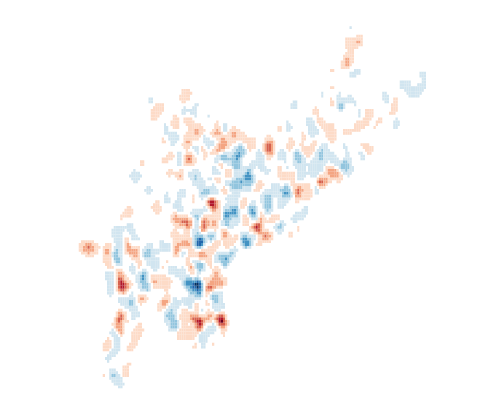}}
 	\fbox{\includegraphics[width=0.22\linewidth]{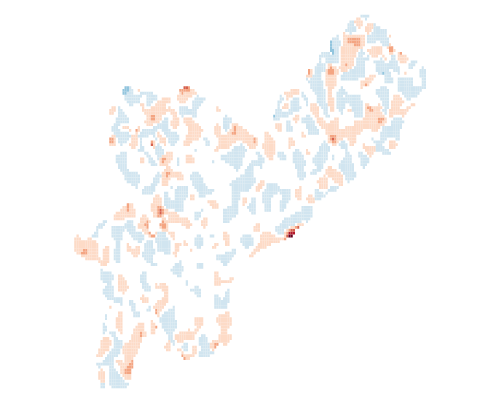}}
 	\fbox{\includegraphics[width=0.22\linewidth]{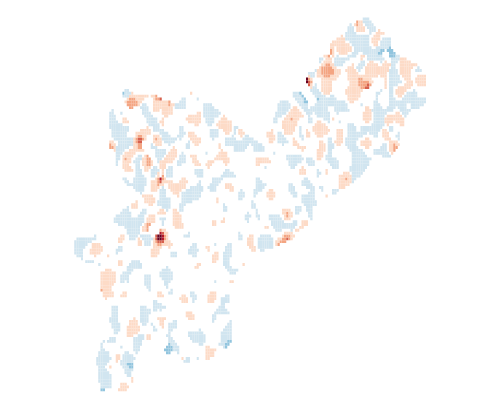}}
 	\vspace{2mm}
 	\\
    \rotatebox{90}{\hspace{11mm} \textsf{Synthetic}} 
 	\fbox{\includegraphics[width=0.22\linewidth]{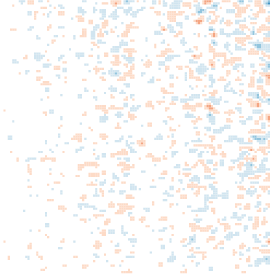}}
 	\fbox{\includegraphics[width=0.22\linewidth]{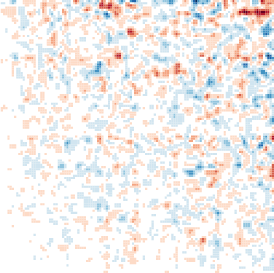}}
 	\fbox{\includegraphics[width=0.22\linewidth]{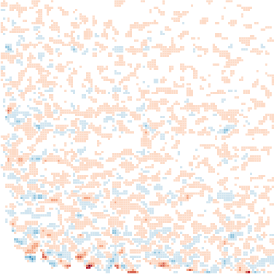}}
 	\fbox{\includegraphics[width=0.22\linewidth]{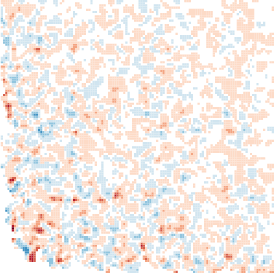}}

  	\includegraphics[width=0.22\linewidth]{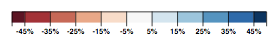}\\
 	
 	\caption{Comparison of the differences between original $\kde$s and coreset $\kde$s (first column) and the difference between original $\kde$s and random sampling $\kde$s (second column).  The last two columns show the corresponding relative differences.}
 	\label{fig:rsvscorediff}
 	\vspace{-5mm}
 \end{figure*}

Another way to understand the error is by directly plotting the error values, as we have done in Figure \ref{fig:rsvscorediff} for the same dataset.  We plot both the absolute and relative error. Here the transfer function is normalized based on the largest difference observed for each dataset and error measure, but held the same between conditions, to allow for the direct comparison of coreset error and random sample error. The resulting color scale is a diverging color map: when the coreset or random sample has a larger value than the true dataset, the area is shown in increasingly saturated shades of red; and when the true dataset has a smaller value, the area is shown in increasingly saturated blue.  When they are similar white is shown.  
We can visually observe darker colors (and hence more error) for the random sampling approach than the coreset approach.  

Note that the theory specifically guarantees the additive error should be smaller for coresets, but we plotted the relative error as well since it seems that such relative differences may have more effect both in quantitative anomaly detection as well as in an observed visual artifact.  Indeed we observe larger relative error for random sampling as well.


%
%



\section{Avoiding Error when Thresholding Iso-Levels}
\label{sec:enet}
A common pattern for interactive data visualization is to show an overview of all the data and then enable analysts to zoom in to investigate regions of interest. For geospatial data, nano-cubes is a recent system that delivers such an experience~\cite{nanocubes} for large datasets.  

A critical aspect of such overviews is hence that they faithfully represent the data in any region above some density of interest, i.e., that wherever there is data above a threshold there should be a visible mark that can be investigated in detail.  
In fact there is a well-developed theory around random sampling regarding this property called and $\eps$-net.  It says if we sample $O((1/\eps) \log (1/\eps))$ points, then any geometric region (like a circle or rectangle) with more than $\eps$-fraction of the points (a density value larger than $\eps$) will contain at least one point~\cite{HW87}.  

However, this desire to show \textbf{all} possibly interesting features runs into another problem.  If we set the minimum threshold for coloring pixels as non-white too low, then the visualization ends up displaying a lot of noise. That is, there may be regions which should have low (or almost $0$) density, which are shown with a visible mark. In contrast to the other sampling results mentioned above (which require larger, $O(1/\eps^2)$-size, samples), the guarantees for $\eps$-nets provide no protection against false positives.  Moreover, simple random sampling is used heavily in many big data systems, such as STORM~\cite{STORM}. 

To address this problem, we will build on a more recent adaption of $\eps$-nets specific to kernel density estimates, called \emph{$(\tau,\eps)$-nets}~\cite{phillips2015subsampling}.  
This coreset $Q \subset P$ ensures that for any point $x \in \b{R}^d$ such that $\kde_P(x) \geq \eps$, there exists a point $q \in Q$ such that $K(x,q) \geq \tau$.  That is, for any query point $x$ above some density threshold $\eps$, there is some \emph{witness} point in the coreset point $q \in Q$ that is nearby (its similarity $K(x,q)$, is at least $\tau$).  
Although such guarantees can be derived from the coresets we discussed earlier, this $(\tau,\eps)$-net only requires a random sample of size $O(\frac{1}{\eps-\tau} \log \frac{1}{\eps-\tau})$, which for $\tau = \eps/2$ is $O(\frac{1}{\eps} \log \frac{1}{\eps})$, i.e., it is roughly the same as the previous and slightly more complex coreset.

So how can we use this idea of a $(\tau, \eps)$-net to aid in choosing a color threshold of our transfer function? One approach is to make that threshold adaptive.  Our proposed method will only color low-density regions (at some threshold taking the place of $\tau$) if they are close to some higher density region (defined by another parameter $\eps$).  This means spurious regions far from the main data will not be illustrated as they are likely noise.  But near a high density region our visualization will draw the lowest density layer.  Data near a high-density regions is less likely to be noise, and so our method displays this part as accurately as possible.  

In detail, we implement this using two values.  The first value $\eps$ (= percentage) is the minimum observed value to represent a ``high density region.''  The second value $r$ (= radius) is the minimum distance an interesting point must be to a high-density region.  Then if a pixel $x$ is not within a distance $r$ of some other pixel $y$ such that $\kde_Q(y) \geq \eps$, then it is not drawn, as if there is no appreciable density there.  If $\kde_Q(x) \geq \eps$ or if $x$ is within distance $r$ of some pixel $y$ such that $\kde_Q(y) \geq \eps$, then it will be drawn as specified by the transfer function.  

\begin{figure*}[h!]

   \hspace{23mm}  \textsf{Full Data} \hspace{40mm}  \textsf{Random Sample}   \hspace{25mm}  \textsf{Random Sample after denoise} \\
 	\rotatebox{90}{\hspace{15mm} \textsf{Kentucky}} 
 	\fbox{\includegraphics[width=0.30\linewidth]{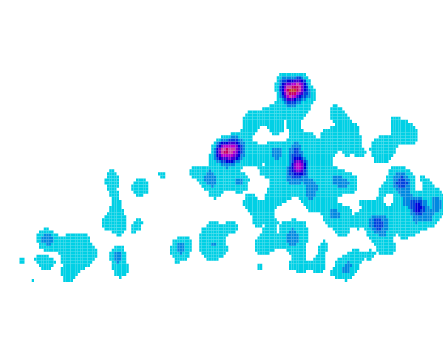}}
 	\fbox{\includegraphics[width=0.30\linewidth]{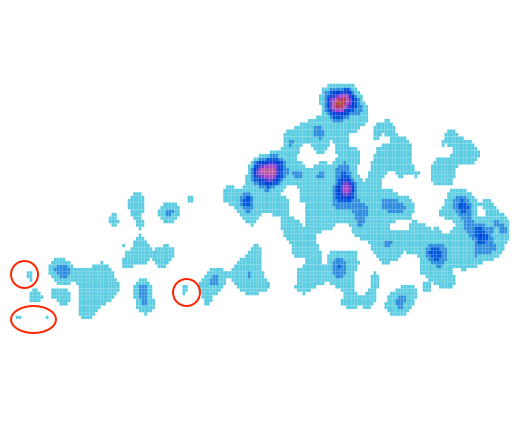}}
 	\fbox{\includegraphics[width=0.30\linewidth]{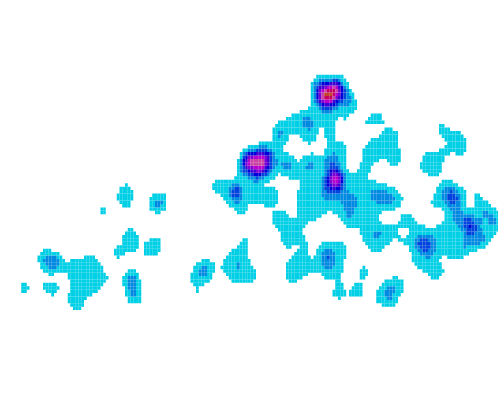}}
 	\rotatebox{90}{\hspace{12mm} \textsf{Philadelphia }} 
 	\fbox{\includegraphics[width=0.30\linewidth]{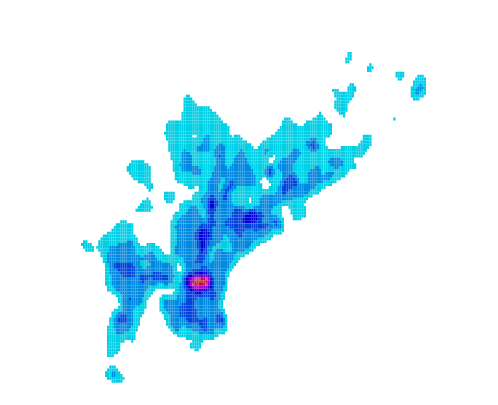}}
 	\fbox{\includegraphics[width=0.30\linewidth]{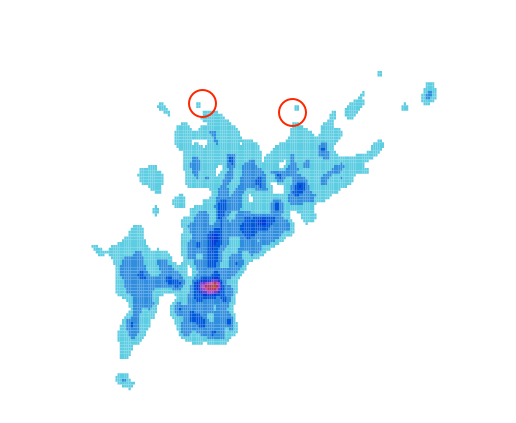}}
 	\fbox{\includegraphics[width=0.30\linewidth]{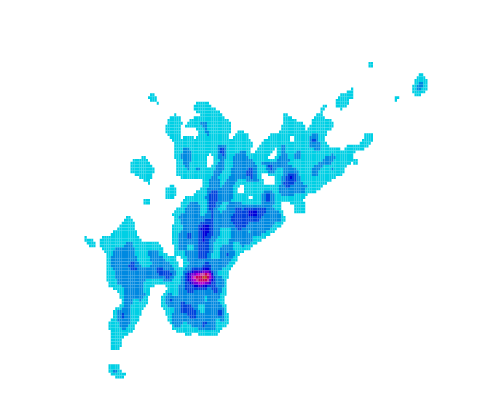}}
    \rotatebox{90}{\hspace{12mm} \textsf{Synthetic}} 
    \fbox{\includegraphics[width=0.30\linewidth]{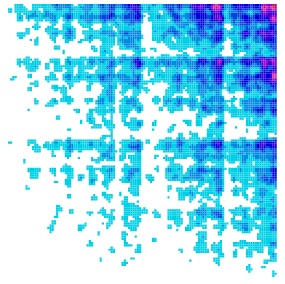}}
 	\fbox{\includegraphics[width=0.30\linewidth]{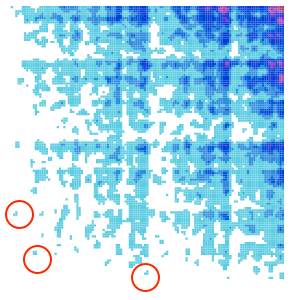}}
 	\fbox{\includegraphics[width=0.30\linewidth]{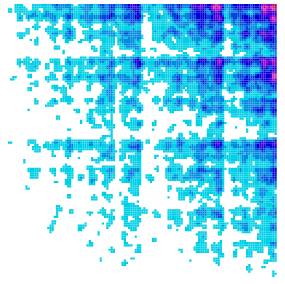}}
 	  \hspace{20mm}  \includegraphics[width=0.25\linewidth]{figs/colorbar1.jpg}\\
 	
 	\vspace{-5mm}
 	\caption{Visualization of random sample $\kde$s of all three datasets.  Showing all isolevels of a random sample (middle) shows false anomalous regions, circled, compared to ground truth (left).  After zapping process, (right) still preserves the rough shape of the data--enough to know where to explore more--without any of the false positive regions. }
 	\label{fig:epsnet}
 	\vspace{-5mm}
 \end{figure*}

Figure \ref{fig:epsnet} demonstrates this approach on our three datasets.  For each dataset, it shows the kernel density estimate for the full data, a random sample of that data, and a de-noised version of the random sample. In the random sample, some anomalous regions appear due to sampling noise (examples are highlighted with red circles in Figure~\ref{fig:epsnet}), which disappear in the de-noised version. The denoised version is a more accurate representation of the original data as it does not show various anomalous bumps of density.

\begin{figure}[t!]
	\includegraphics[width=0.98\linewidth]{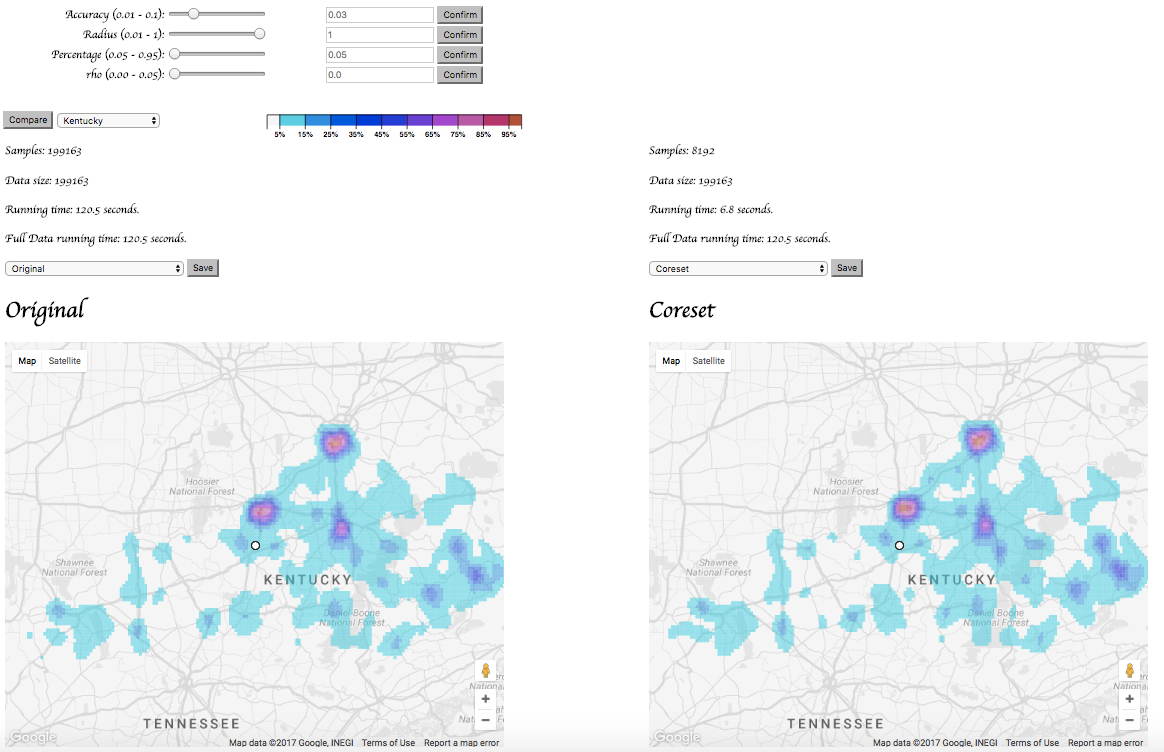}
	\caption{A snapshot of the system.}
	\label{fig:system}
\end{figure}

\section{System}
To demonstrate our approach and compare it to both, ground truth and random sampling we build an interactive system to display kernel density estimates of very large spatial data. It enables analysts to interactively explore such large data while avoiding false positives. To enable a direct comparison of various approaches, we show two windows showing the same dataset using different methods---the $\kde$ of the full dataset, coreset $\kde$, random sample $\kde$, coreset error, coreset relative error, random sampling error and random sampling relative error. Analysts can specify the error threshold $\eps$, based on which the system automatically generates a coreset or a random sample based on $\eps$.

Zooming and panning is synchronized between views, so that analysts can navigate and compare the views at various scales and positions. To provide geospatial context, the $\kde$ visualization is rendered on top of a customized Google Map widget, which shows the geographic features as grayscale to avoid interference with the colors used to display the $\kde$. 

\begin{figure}[t!]
 \includegraphics[width=0.49\linewidth]{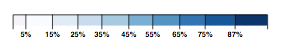}
  \includegraphics[width=0.49\linewidth]{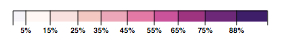}
  \includegraphics[width=0.49\linewidth]{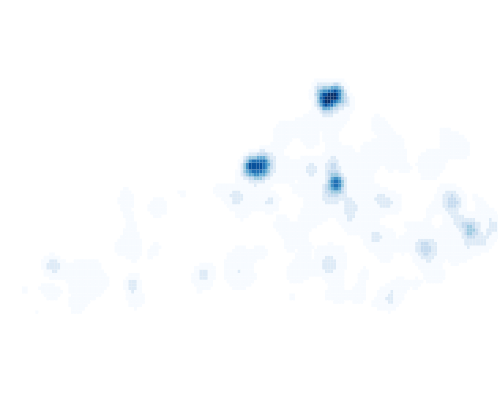}
  \includegraphics[width=0.49\linewidth]{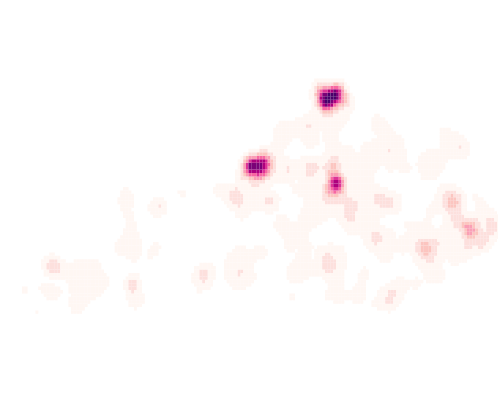}
   	\vspace{-5mm}
  \caption{Visualization of $\kde$s with different colorbars.}
   \label{fig:changecolor}
\end{figure}

We also provide various color maps options from ColorBrewer \cite{colorbrewer}.   We allow users dynamically change the choice of color map, and its scaling within the colorbar (Figure \ref{fig:changecolor}).

\subsection{Interactive De-noising}

\begin{figure}[t]
\begin{centering}
\subfigure[$\kde$ of a random sampling and selected zapping area.]{
\includegraphics[width=0.98\linewidth]{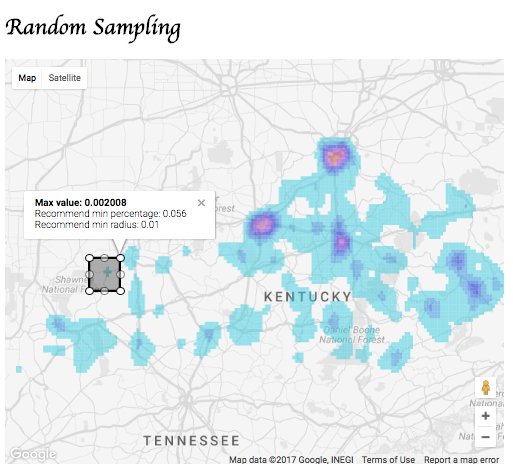}
\label{fig:zap1}
}
\subfigure[Input the selected parameters.]{
\includegraphics[width=0.98\linewidth]{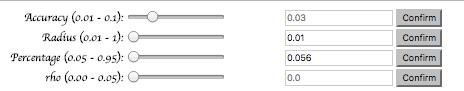}
\label{fig:zap2}
}
\caption{Illustration of the interactive de-noising process. Analysts select a region in the visualization they suspect to contain an artifact. The algorithm suggests parameters that can be used to remove that artifact (a) and applies them to the input fields (b).}
\label{fig:zapping}
\end{centering}
\end{figure}

When applying the de-noising process that alters the low end of the color scale with $\eps,\tau$-nets, we found that the choice of these parameters can be difficult for a user to select. To address this, we designed a feature where an analyst can highlight a region that appears to be an anomalous region, and the system will suggest the a pair of minimal percentage and radius values that can be set to remove the noise in that region.  Figure \ref{fig:zapping} illustrates this process within our system.  

Analysts select an isolated regions to get rid of, then a tips message will give the suggestions of ``percentage" and ``radius",  so $\tau =$ ``percentage" $\times$ the largest $\kde$ within ``radius" of the objective point. These values can then be applied to the parametrization of the de-noising process, eliminating the noisy spot and other like it. 

We suggest to users to attempt this with a few isolated dots and see the effects.  Then if desired, they can also manually tune these parameters directly and quickly see the effect.  

\subsection{Implementation}

The front end of our technology demonstration is implemented in HTML/JavaScript and uses D3 for axis, scales and user interface elements, Canvas for the rendering of the KDEs and the Google Maps API for the background maps.

The backend that generates the coresets is an extension of work from SIGMOD 2012~\cite{big-kde} and is written in C.  This can take any large spatial dataset as a text file, a error parameter $\eps$, and output a coreset.  We modify this to generate a priority reordering of the entire dataset so that every initial subset of the data is a coreset, with error parameter effectively decreasing as the chosen coreset size increases.  This process is also written in C, and generates a text file sufficient for the HTML/Javascript to use as its input.  

The implementation of the visualization system (\url{https://github.com/SayingsOlly/kernel_vis_d3}) and the back-end code (\url{http://www.cs.utah.edu/~yanzheng/kde/}) is available under the BSD 3-clause license.   We invite others to download and interact with it.

\section{Discussion and Limitations}

We study the specific but ubiquitous visualization tool of kernel density estimates, with the goal of how best to integrate them into a large-scale visualization system --- specifically those making the increasingly common design choice to approximate massive datasets.  In this context we demonstrate that coresets provide better and more efficient estimates than simple random sampling.  We also develop a new way to preprocess the coresets so that their size resolution can be easily updated without redoing expensive computation.  
Additionally, we introduced a new tool for dealing with spatial noise at low densities --- a common nuisance that distracts the user to explore potential outliers which are not present in the full dataset.  This provides an easy way to ``zap'' these unfortunate event with a simple rule that will apply to all similar visual (but not statistical) anomalies.  
Our simple system demonstrates the usefulness of all of these insights through interaction with real and synthetic dataset.

Our interactive visualization system, however, is designed as a prototype to demonstrate the strengths of the underlying technique and is not designed to be a fully-fledged geospatial data analysis system. Several improvements with respect to data loading and usability are conceivable to make the system useful for actual analysis tasks. 
We would also like to explore the effects of different coreset constructions (e.g., \cite{Phillips2013,CWS10}) and types of kernels other than Gaussians (e.g., Laplace or Epanechnikov).  

With any interactive visualization tool, it is important to be cognizant of the potential for \emph{visual p-hacking}~\cite{WCHB10}:  where a user tweaks the visual parameters until he/she finds the interpretation of the data he/she wants to see, but unwittingly has just discovered artifacts of the noise in the data.  Our technique moderates this by allowing users to identify noise (perhaps using expert knowledge) and remove it.  Moreover, it enforces the same pruning criteria for all isolated parts of the dataset, so it is not possible to design pruning criteria separately for different areas --- an easy way to overfit.  

In general, one should compliment this with a query-and-filter strategy to verify abnoromal or interesting aspects of the data beyond just the visual patterns.  Our tool is meant to help users quickly determine where to take these closer looks.  


\section{Conclusion and Future Work}

We have demonstrated the use of coresets for kernel density estimates, ways to preprocess them for easy parameter updates, and how to prune a certain type of low-density noise.  We believe these are techniques that should be integrated into many visualization systems for large spatial datasets.
  
However, our system itself is only a prototype.  We would like to actually map these ideas into more complex systems (e.g., nanocubes~\cite{nanocubes} or STORM~\cite{STORM}) which already deal with and approximate various datasets and allow for other richer types of interactions.  

We also believe coresets~\cite{Phi16,BLK17} can potentially be a very useful tool for efficiently visually interacting with many types of massive datasets.  We hope to explore more of these applications in the future.

\acknowledgements{
Thanks to support by NSF CCF-1350888, IIS-1251019, ACI-1443046, CNS-1514520, CNS-1564287 and NIH U01 CA198935.}

\bibliographystyle{abbrv}
\bibliography{kernel-refs}
\end{document}